\documentclass[pdflatex,11pt]{article}
\usepackage{graphicx} 
\usepackage{amsmath,amssymb,amsfonts,bm,cancel} 

\def\a{\alpha} \def\b{\beta}   \def\d{\delta}  \def\e{\epsilon}  \def\h{\eta} \def\th{\theta}    \def\L{\Lambda} \def\m{\mu} \def\n{\nu}   \def\p{\pi}  \def\r{\rho} \def\s{\sigma}  \def\t{\tau}       

\def\dg{\dagger}  \def\nn{\nonumber}

\newcommand{\lsp}{ \left ( } \newcommand{\rsp}{ \right ) }   \newcommand{\To}{\Rightarrow}   

\newcommand{\vev}[1]{ \langle {#1} \rangle }

 \renewcommand{\Im}{{\rm Im \, }}

\newcommand{\diag}[2]{ \begin{pmatrix}  #1 & 0 \\ 0 & #2 \\   \end{pmatrix}  }

\newcommand{\Diag}[3]{ \begin{pmatrix} #1 & 0 & 0 \\ 0 & #2 & 0 \\ 0 & 0 & #3 \\\end{pmatrix}}

\begin{document}

\begin{flushright}
STUPP-24-271
\end{flushright}

\vskip 1.35cm

\begin{center}

{\large \bf Almost general analysis of CKM and MNS matrices 
 for hierarchical Yukawa structure and interpretation of Dirac CP phase probed by DUNE and T2HK }

\vskip 1.2cm

Masaki J. S. Yang

\vskip 0.4cm

{\it Department of Physics, Saitama University, \\
Shimo-okubo, Sakura-ku, Saitama, 338-8570, Japan\\
}



\begin{abstract} 

In this letter, we perform an almost general analysis of flavor-mixing matrices $V_{\rm CKM}$ and $U_{\rm MNS}$ to investigate the discriminative power of CP phases by next-generation neutrino oscillation experiments.
As an approximation, we neglect the 1-3 mixing of diagonalization for 
more hierarchical fermions $u,e$. 
Thus there are two sources of CP violation in $V_{\rm CKM}$ and $U_{\rm MNS}$,
 the intrinsic CP phase $\delta_{d, \nu}$ in diagonalization of less hierarchical fermions $d, \nu$ and relative phases between two unitary matrices. 
By eliminating unphysical phases and imposing constraints of the three measured mixing angles, 
the flavor-mixing matrices are analytically displayed by two phases and the 1-2 mixing $s_{u, e}$ of more hierarchical fermions.  
For sufficiently small 1-2 mixing $s_{e}$ of charged leptons, 
the Dirac phase $\delta$ is mostly identical to the intrinsic phase of neutrinos $\delta_{\nu}$. 
Therefore, future detection of the Dirac phase indicates the observation of $\delta_{\nu}$. 
On the other hand, if such a CP violation is not observed, 
an upper limit is placed on a combination of $\delta_{\nu}$ and relative phases.

\end{abstract} 

\end{center}

\section{Introduction}

The CP violation (CPV) in the lepton sector is one of the important physical quantities that should be measured precisely in the next decade. 
The Dirac phase $\d$ representing this violation has been searched by T2K experiment \cite{T2K:2021xwb} and NO$\n$A experiment \cite{NOvA:2021nfi} respectively.
There are significant discrepancies between the two measurements, and 
this tension is expected to be alleviated or resolved by next-generation neutrino oscillation experiments. 
Tokai to Hyper-Kamiokande (T2HK) Experiment \cite{Hyper-KamiokandeProto-:2015xww} and 
Deep Underground Neutrino Experiment (DUNE) \cite{DUNE:2020jqi} have the ability to find the Dirac phase of around 30$^{\circ}$ after 10 years of operation. 

In this current observational circumstance, a question arises; 
what kind of restrictions are generally placed on CP phases in the lepton sector, when $\d$ is or is not observed by these experiments? 
For the CKM matrix, general analyses exist for hierarchical Yukawa matrices \cite{Hall:1993ni, Peccei:1995fg} and many flavor texture studies \cite{Xing:2020ijf}.
Phenomenological analyses have also been studied for the next-generation neutrino oscillation experiments \cite{C:2014mmz, Ballett:2016daj, deGouvea:2017yvn,Kelly:2017kch} and 
the Dirac phase \cite{Ge:2011qn, Ge:2011ih, Petcov:2014laa, Girardi:2014faa, Girardi:2015vha, Delgadillo:2018tza}. 
However, no general treatment for each unitary matrix has been found for the two flavor-mixing matrices $V_{\rm CKM} = U_{u}^{\dg} U_{d}$ and $U_{\rm MNS} = U_{e}^{\dg} U_{\n}$.
Therefore, in this letter, we analyze the mixing matrices of quarks and leptons in an almost general form 
for hierarchical mass structures of the fermions 
and discuss the distinguishability of CP phases of leptons by the next-generation experiments. 

%
%

\section{An almost general representation of flavor-mixing matrices for hierarchical Yukawa structure}

This section introduces an almost general form of flavor-mixing matrices 
 that originate from hierarchical Yukawa matrices.
At first, we define that a Yukawa matrix $Y_{f}$ of the Standard Model fermions $f = u,d,\n, e$ is {\it hierarchical} if its elements satisfy $|Y_{f ii}| \gg |Y_{f i j}| , |Y_{ji}|$ for $i > j$. 
Namely, for a matrix, 
\begin{align}
Y_{f} \equiv 
\begin{pmatrix}
Y_{f11} & Y_{f12} & Y_{f13} \\
Y_{f21} & Y_{f22} & Y_{f23} \\
Y_{f31} & Y_{f32} & Y_{f33} \\
\end{pmatrix} , 
\end{align}
absolute values of its elements $|Y_{fij}|$ satisfy
\begin{align}
|Y_{f33}| \gg |Y_{f 23}| , |Y_{f 32}| , |Y_{f13}| , |Y_{f31}| \, , 
~ {\rm and} ~ 
|Y_{f22}| \gg |Y_{f 21}| , |Y_{f 12}| \, . 
\end{align}
With these conditions, 
unitary matrices of the singular value decomposition of $Y_{f}$ necessarily has only small mixings,  
and singular values $m_{fi}$ of $Y_{f}$ are close to its diagonal elements $|Y_{f ii}| \simeq m_{fi}$.

These hierarchical Yukawa matrices have approximate chiral flavor symmetries $U(1)_{L}^{2} \times U(1)_{R}^{2}$; 
\begin{align}
Y_{f}' 
= \Diag{e^{i \a_{1 L}}}{e^{i \a_{2 L}}}{1} 
Y_{f} 
\Diag{e^{i \a_{1 R}}}{e^{i \a_{2 R}}}{1} 
\simeq Y_{f} \, . 
\end{align}
Such $Y_{f}$ is easily realized by hierarchical vacuum expectation values (vevs) of scalar (flavon) fields with appropriate chiral charges.
Specifically, we consider a situation where
 the Yukawa matrices are generated by vevs  of flavon fields $\phi_{i}$ and $\tilde \phi_{j}$ that have chiral charges associated with the left-handed $i$ and right-handed $j$ generations. 
 At leading order, Yukawa matrices are generated as 
\begin{align}
Y_{f} = {1\over \Lambda^{2}}
\begin{pmatrix}
y_{f11} \vev{\phi_{1}} \vev{\tilde \phi_{1}} & y_{f12} \vev{\phi_{1}} \vev{\tilde \phi_{2}} & y_{f13}  \vev{\phi_{1}} \vev{\tilde \phi_{3}} \\
y_{f21}  \vev{\phi_{2}} \vev{\tilde \phi_{1}} & y_{f22}  \vev{\phi_{2}} \vev{\tilde \phi_{2}} & y_{f23}  \vev{\phi_{2}} \vev{\tilde \phi_{3}} \\
y_{f31} \vev{\phi_{3}} \vev{\tilde \phi_{1}}  & y_{f32} \vev{\phi_{3}} \vev{\tilde \phi_{2}}  & y_{f33} \vev{\phi_{3}} \vev{\tilde \phi_{3}}  \\
\end{pmatrix} .
\end{align}
Here, $y_{fij}$ denotes coupling constants between flavons and fermions $f$, and $\L$ is some cutoff scale. Approximate chiral symmetries exist if 
these vevs satisfy $|\vev{\phi_{3}}| \gg |\vev{\phi_{2}}| \gg |\vev{\phi_{1}}|$ and 
$|\vev{\tilde \phi_{3}}| \gg |\vev{\tilde \phi_{2}}| \gg |\vev{\tilde \phi_{1}}|$. 
Various models of mass matrices exhibit these properties.

\subsection{An approximation from hierarchy of Yukawa matrices}

For the hierarchical Yukawa matrices, 
this letter deals with two flavor-mixing matrices, 
the CKM matrix $V_{\rm CKM}$ and the MNS matrix $U_{\rm MNS}$.
\begin{align}
V_{\rm CKM} = U_{u}^{\dg} U_{d} \, , ~~~
U_{\rm MNS} = U_{e}^{\dg} U_{\n} \, . 
\end{align}
Here, $U_{f}$ denote left-handed unitary matrices that diagonalize $Y_{f}$ and the mass matrix of  neutrinos $m_{\n}$.
This $U_{f}$ is generally displayed by the PDG parameterization as 
\begin{align}
U_{f} = \Phi_{L}^{f} U_{f}^{0} \Phi_{R}^{f} \, , 
\end{align}
with diagonal phase matrices $\Phi^{f}_{L,R}$. 
The matrix $U_{f}^{0}$ is a function $U_{f}^{0} = f (s_{12}^{f}, s_{23}^{f}, s_{13}^{f}, \d^{f})$ of mixing angles $s_{ij}^{f}$ and  the intrinsic CP phase $\d^{f}$. 

To make the analysis concise, the following approximation is imposed.
\begin{description}
\item[Approximation:] We neglect the 1-3 mixing of $U_{f}$ 
for fermions $f$ with a stronger hierarchy of singular values (i.e. $f = u ,e$). 

\item[Justification:] The 1-3 mixing of $U_{f}$ is suppressed by  $m_{f1}/m_{f3}$ (or a power of the ratio) by the approximate chiral symmetry described above. 
\end{description}
To be specific, if the unitary matrices diagonalizing $Y_{f}$ have small mixings, 
approximate chiral symmetries roughly constrain the texture of $Y_{f}$ as
\begin{align}
Y_{f} \simeq (Y_{f})_{33}
\begin{pmatrix}
\e_{L}^{f} \e_{R}^{f} & \e_{L}^{f} \d_{R}^{f} & \e_{L}^{f} \\
\d_{L}^{f} \e_{R}^{f}  & \d_{L}^{f} \d_{R}^{f} & \d_{L}^{f} \\
\e_{R}^{f} & \d_{R}^{f} & 1
\end{pmatrix} , 
\end{align}
where $\e_{L,R}^{f}, \d_{L,R}^{f}$ are breaking parameters of the $U(1)_{(L,R)}$ symmetry for the first and second generations, and $O(1)$ coefficients are omitted.
The behavior of the 1-3 mixing is divided into three cases depending on the magnitude of $\e_{L,R}^{f}$. 
\begin{enumerate}
\item $|\e_{R}^{f}| \sim 1 \gg |\e_{L}^{f}|$: 
From $(Y_{f})_{11} / (Y_{f})_{33} \simeq m_{f1}/ m_{f3}$, 
the 1-3 mixings are approximately evaluated as 
\begin{align}
|s_{13}^{e}| & \simeq |\e_{L}| \simeq m_{e1} / m_{e3} \simeq 0.0003 \, , \\
|s_{13}^{u}| & \simeq |\e_{L}| \simeq m_{u1} / m_{u3} \simeq 0.00001 \, . 
\end{align}
These mixings are safely neglected because they are sufficiently smaller than the 1-3 elements of the flavor-mixing matrices $|(V_{\rm CKM})_{13}| \sim 10^{-3}, |(U_{\rm MNS})_{13}| \sim 10^{-2}$.  
Besides, for too large $\e_{R}^{f} \gtrsim 0.1$, there is no longer the approximate symmetry for right-handed fields.

\item $|\e_{L}^{f}| \sim |\e_{R}^{f}|$: 
Similarly, 
\begin{align}
|s_{13}^{e}| & \simeq |\e_{L}| \simeq \sqrt{m_{e1} / m_{e3}} \simeq 0.017 \, , \\
|s_{13}^{u}| & \simeq |\e_{L}| \simeq \sqrt{m_{u1} / m_{u3}} \simeq 0.0027 \, .
\end{align}
Its contribution for leptons is at most a $\pm$10\% change in $U_{e3}$. 
On the other hand, the 1-3 mixing for quarks can have a large impact on $|V_{ub}| \simeq 0.003$. However,  a similar hierarchy for down-type quarks leads to $ \sqrt{m_{d1} / m_{d3}} \simeq 0.03$, and such a model with large 1-3 mixings is excluded unless a fine-tuned cancellation.

\item $|\e_{L}| \gg |\e_{R}|$: 
In this case, it is also hard to consider that there is an approximate $U(1)_{L}$ symmetry.
The approximation does not hold and the unitary matrix for left-handed fields has large mixings.  
With the corresponding CKM matrix element $|V_{ub}| \simeq 0.003$ in mind, 
this situation seems difficult to realize in a unified theory. 

\end{enumerate}
Thus, this approximation is justified in a wide range of parameters.

A similar discussion allows some mention for magnitudes of the 1-2 mixings $\e_{L}^{f} / \d_{L}^{f}$.
\begin{align}
&{m_{e1} \over m_{e2}} \simeq  0.005  \lesssim |s_{12}^{e}|  \lesssim 0.07
 \simeq   \sqrt{m_{e1} \over m_{e2}} \, ,  \\
& {m_{u1} \over m_{u2}} \simeq 0.002  \lesssim |s_{12}^{u}| \lesssim 0.04
 \simeq \sqrt{m_{u1} \over m_{u2}}  \, . 
\end{align}
In subsequent analyses, we consider parameter regions of $s_{12}^{e, u} \lesssim 0.1$.

\subsection{Representation of flavor-mixing matrices}

Using this approximation, we organize an almost general representation of the flavor-mixing matrices. 
For unitary matrices $U_{\n, e} = \Phi_{L}^{\n , e} U_{\n ,e}^{0} \Phi_{R}^{\n, e}$ displayed in the PDG parameterization, 
there is relative phases $\Phi_{L}^{e \dg} \Phi_{L}^{\n} = $ diag $(e^{i \r}, e^{i\s}, e^{i \s '})$ between $U_{\n}^{0}$ and $U_{e}^{0}$. 
By a redefinition of the overall phase, 
we define $\s' \equiv - \s$ to set the determinant of the 2-3 submatrix to unity.
Then the MNS matrix is written by
\begin{align}
U_{\rm MNS} &\equiv \Phi_{R}^{e \dg}
U_{e}^{0 \dagger}
 \Diag{e^{i \r}}{e^{i \s}} {e^{-i\s}}
U_{\n}^{0} \Phi_{R}^{\n},  ~~~ 
U_{e}^{0 \dagger} \equiv
\begin{pmatrix}
c_{e} & - s_{e} & 0 \\
s_{e} & c_{e} & 0 \\
 0 & 0 & 1 \\
\end{pmatrix} 
\begin{pmatrix}
 1 & 0 & 0 \\
 0 & c_{\t} & -  s_{\t} \\
 0 & s_{\t} & c_{\t} \\
\end{pmatrix} \, , \nn \\
U_{\n}^{0} &\equiv
\begin{pmatrix}
 1 & 0 & 0 \\
 0 & c_{\n} & s_{\n} \\
 0 & - s_{\n} & c_{\n} \\
\end{pmatrix}
\begin{pmatrix}
c_{13} & 0 & s_{13} e^{-i \d_{\n}} \\
 0 & 1 & 0 \\
- s_{13} e^{i \d_{\n}} & 0 & c_{13} \\
\end{pmatrix}
\begin{pmatrix}
c_{12} & s_{12} & 0 \\
 - s_{12} & c_{12} & 0 \\
 0 & 0 & 1 \\
\end{pmatrix}  . 
\label{UVn}
\end{align}
Here $s_{f} \equiv \sin f\, , \, c_{f} \equiv \cos f \, , \, s_{ij} \equiv \sin \th_{ij} \, , \, c_{ij} \equiv \cos \th_{ij}$. 
Since the phases $\Phi_{R}^{\n, e}$ do not affect the Dirac phase, they are omitted hereafter. 
The CKM matrix is also defined in the same way. 

The approximation eliminates the intrinsic CP phase $\d_{u,e}$ associated with $U_{u,e}$, 
because three generations are no longer involved in flavor-mixing for the more hierarchical fermions.
Thus, in this analysis, there are two sources of CP violation: 
relative phases between the two unitary matrices $\r, \s$ and 
intrinsic CP phase $\d_{d, \n}$ for the less hierarchical fermions.

The two 2-3 mixings are merged to define new arguments $\chi$ and $\h$.
\begin{align}
& 
\begin{pmatrix}
 c_{\t} & - s_{\t} \\
 s_{\t} & c_{\t} \\
\end{pmatrix} 
\diag{e^{i \s}}{e^{ - i \s}}
\begin{pmatrix}
 c_{\n} & s_{\n} \\
 - s_{\n} & c_{\n} \\
\end{pmatrix}
= 
\begin{pmatrix}
e^{i \s} c_{\t} c_{\n} + e^{-i\s} s_{\t} s_{\n} &e^{i \s} c_{\t} s_{\n} - e^{-i\s} s_{\t} c_{\n}  \\
e^{i \s} s_{\t} c_{\n} -  e^{-i\s} c_{\t} s_{\n} & e^{i \s} s_{\t} s_{\n} +  e^{-i\s} c_{\t} c_{\n} \\
\end{pmatrix} 
\\ &  = 
\begin{pmatrix}
c_{\s} c_{\n - \t} + i s_{\s} c_{\n + \t} &  c_{\s} s_{\n - \t} + i  s_{\s} s_{\n + \t}  \\
 - c_{\s} s_{\n - \t} + i s_{\s} s_{\n + \t} & c_{\s} c_{\n - \t} - i s_{\s} c_{\n + \t}  \\
\end{pmatrix}
\equiv 
\begin{pmatrix}
e^{ i \chi} c_{23} & e^{i \h} s_{23} \\
 - e^{- i \h} s_{23} & e^{- i \chi} c_{23} \\
\end{pmatrix} \, . 
\end{align}
Relations between these arguments $\chi, \h$ and parameters $\s, \n, \t$ are
\begin{align}
\tan \chi = \tan \s { c_{\n + \t} \over  c_{\n - \t}} \, ,  ~~~ 
\tan \h = \tan \s {s_{\n + \t} \over s_{\n - \t}} \, . 
\end{align}
From this, 
\begin{align}
U_{\rm MNS} & =
\begin{pmatrix}
c_{e} & - s_{e} & 0 \\
s_{e} & c_{e} & 0 \\
 0 & 0 & 1 \\
\end{pmatrix} 
\begin{pmatrix}
 e^{i \r} & 0 & 0 \\
0 & e^{ i \chi} c_{23} &  e^{i \h} s_{23} \\
0 &- e^{- i \h} s_{23} & e^{- i \chi} c_{23} \\
\end{pmatrix}
\begin{pmatrix}
c_{13} & 0 & s_{13} e^{-i \d_{\n}} \\
 0 & 1 & 0 \\
- s_{13} e^{i \d_{\n}} & 0 & c_{13} \\
\end{pmatrix}
\begin{pmatrix}
c_{12} & s_{12} & 0 \\
 - s_{12} & c_{12} & 0 \\
 0 & 0 & 1 \\
\end{pmatrix} \, , 
\end{align}
Furthermore, by multiplying diag $(1 , 1, e^{i (\chi + \h)})$ from the left and diag $(1,1,e^{ i (\chi - \h)})$ from the right of $U_{\rm MNS}$ and removing an overall phase $e ^{ i \chi}$, 
we obtain
\begin{align}
U_{\rm MNS} & =
\begin{pmatrix}
c_{e} & - s_{e} & 0 \\
s_{e} & c_{e} & 0 \\
 0 & 0 & 1 \\
\end{pmatrix} 
\begin{pmatrix}
 e^{i (\r - \chi)} & 0 & 0 \\
0 & c_{23} & s_{23} \\
0 & - s_{23} & c_{23} \\
\end{pmatrix}
\begin{pmatrix}
c_{13} & 0 & s_{13} e^{-i (\d_{\n} + \h - \chi)} \\
 0 & 1 & 0 \\
- s_{13} e^{i (\d_{\n} + \h - \chi)} & 0 & c_{13} \\
\end{pmatrix}
\begin{pmatrix}
c_{12} & s_{12} & 0 \\
 - s_{12} & c_{12} & 0 \\
 0 & 0 & 1 \\
\end{pmatrix} \, . 
\label{UVn2}
\end{align}
Therefore, the following two phases $\a$ and $\b$ have physical significance.
\begin{align}
\a \equiv \rho - \chi \, , ~~~ \b = \d_{\n} + \h - \chi \, . 
\end{align}
If $s_{\t}$ is sufficiently small by the chiral symmetry, 
$\chi \sim \eta \sim \s$ and $\a \sim \rho - \s \, , \, \b \sim  \d_{\n}$ hold.
This notation is a generalization of the Fritzsch--Xing parameterization \cite{Fritzsch:1997fw}.

Finally, we discuss sign degrees of freedom for mixing angles. 
Continuous parameters in this notation are $s_{e}, s_{ij}$ and two phases $\a, \b$. 
Similar to the PDG parameterization, phase redefinitions of six fermions can eliminate five signs. 
Since the sign of $s_{13}$ can be absorbed into the phase $e^{\pm i \b}$, 
the six $s_{ij}$ and $c_{ij}$ are set to be positive.
By multiplying diag $(-1, -1, 1)$ from the left of Eq.~(\ref{UVn2}), the sign of $c_{e}$ is inverted 
 without changing the sign of $s_{ij}$ and $c_{ij}$.
As a result, only the sign of $s_{e}$ is physical because the sign of two $s_{e}$ cannot be absorbed into the phase $\a$. 
Later, we specifically check that the physical sign of the CP violation is sign$(c_{e} s_{e})$. 

\subsection{Constraints on mixing angles}

The six continuous parameters ($s_{e}, s_{ij}, \a$  and $\b$) are constrained by the three observed mixing angles, resulting in three free parameters $\a, \b, s_{e}$ in the MNS matrix. 
For this purpose, 
the best-fit values of the latest global fit \cite{Gonzalez-Garcia:2021dve} are used as input parameters for the MNS matrix. 
The three mixing angles 
of the normal hierarchy (NH) without Super-Kamiokande (SK) are 
\begin{align}
& \sin^{2} \th_{12}^{\rm NH} = 0.304 \, , ~~ \sin^{2} \th_{23}^{\rm NH} = 0.573 \, , ~~~ \sin^{2} \th_{13}^{\rm NH} = 0.0222 \, . 
\end{align}
The reason for using NH without SK is that the values of inverted hierarchy (IH) with or without SK are close to these values.  
Although the inclusion of the SK data makes $\th_{23}^{\rm NH}$ about $0.1$ smaller, 
the qualitative behavior remains the same in the following discussion.

Three mixing parameters $s_{ij}$ are constrained from $\th_{ij}^{\rm NH}$.
Hereafter, the above three input parameters are denoted as $\sin \th_{ij}^{\rm NH} \equiv S_{ij} \, , \cos \th_{ij}^{\rm NH} \equiv C_{ij}$. 
First, the 3-3 element of $U_{\rm MNS}$, which is the same as the PDG parameterization, determines $s_{23}$; 
\begin{align}
c_{13} c_{23} = C_{13} C_{23} \, , ~~~ 
s_{23} = \sqrt{1 - {C_{13}^{2} C_{23}^{2} \over c_{13}^{2}}} \, . 
\label{sol23}
\end{align}
Next, $|U_{13}|^{2}$ is a function of $s_{13}$ and $s_{23}$. 
\begin{align}
U_{13} & = c_{e} s_{13} e^{i (\a - \b)} - c_{13} s_{23} s_{e} \, ,  \label{U13}\\
|U_{13}|^{2} & = c_{e}^2 s_{13}^2 +  c_{13}^2 s_{23}^2 s_{e}^2 - 2 c_{13} c_{e} s_{13} s_{23} s_{e} \cos(\a - \b) \, . \label{U132}
\end{align}
Substituting the solution of $s_{23}$ and setting $|U_{13}|^{2}$ equal to $S_{13}^{2}$, 
we obtain a quadratic equation for $s_{13}^{2}$.
\begin{align}
& s_{13}^4 (4 c_{e}^2 s_{e}^2 \cos^2(\a - \b)+(s_{e}^2-c_{e}^2)^2 )
+(s_{e}^2 (C_{13}^2 C_{23}^2-1)+S_{13}^2)^2  \nn \\
 & + 
s_{13}^{2} [2 (s_{e}^2-c_{e}^2) (s_{e}^2 (C_{13}^2 C_{23}^2 - 1)+S_{13}^2) + 
4 c_{e}^2 s_{e}^2 (C_{13}^2 C_{23}^2-1) \cos^2(\a-\b)] = 0 \, . 
\end{align}
The two solutions of $s_{13}^{2}$ are
\begin{align}
s_{13}^{2} &= 
{ A \pm B
\over (c_{e}^4 + 2 c_{e}^2 s_{e}^2 \cos 2 (\a - \b) + s_{e}^4)} \, ,  \label{sol13} \\
A & = 
- c_{e}^2 s_{e}^2 (C_{13}^2 C_{23}^2 - 1) \cos 2 (\a - \b) 
-C_{13}^2 C_{23}^2 s_{e}^4 + c_{e}^2 S_{13}^2 - S_{13}^2 s_{e}^2 + s_{e}^4 \, ,  \\
B & =  2 c_{e} s_{e} \cos (\a - \b)
\sqrt { C_{13}^2 S_{13}^2 S_{23}^2 - c_{e}^2 s_{e}^2 (C_{13}^2 C_{23}^2 - 1)^2 \sin^2(\a - \b) } \, . 
\end{align}
For sufficiently small $s_{13}$, the solution $A + B$ is chosen because it reproduces the correct $U_{13}$. 
The other solution corresponds to the dashed line in Fig.~1, 
because the sign is identified with the sign of $\cos(\a -\b)$. 

From the triangular inequality for Eq.~(\ref{U13}), upper and lower bounds exist on $s_{13}$.
\begin{align}
c_{e} S_{13} - C_{13} S_{23} s_{e} \leqq s_{13} \leqq C_{13} S_{23} s_{e} + c_{e} S_{13} \, . 
\end{align}
This result is easy to understand because
the mixing matrix of $s_{ij}$ is a ``reverse rotation by $s_{e}$'' of the PDG representation of $S_{ij}$.
From this, there exists a critical value $s_{e}^{c}$ of $s_{e}$ such that $s_{13} = 0$.
\begin{align}
s_{13} = 0  ~~ \To ~~ {s_{e}^{c} \over c_{e}^{c}} = {S_{13} \over C_{13} S_{23}} \, . 
\end{align}
This value is about 0.2 for the MNS matrix (and 0.09 for the CKM matrix).
For larger $s_{e} > s_{e}^{c}$, as can be seen in Fig.~1, the two solutions in Eq.~(\ref{sol13}) are  physically inequivalent.

\begin{figure}[t]
\begin{center}
 \includegraphics[width=13cm]{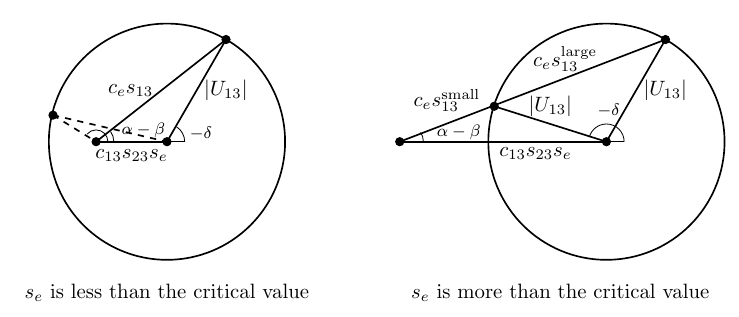}
\caption{Geometric explanation of solutions of $s_{13}$ for given mixing angles $s_{23}$ and $s_{e}$. 
If $s_{e}$ is below the critical value, there are two solutions with different signs of $\cos (\a - \b)$.
If $s_{e}$ is above the critical value, there are two solutions with different magnitudes
 for the same $\cos (\a - \b)$.
}
\end{center}
\end{figure}

Finally, we consider the condition on $s_{12}$. 
The matrix element $U_{12}$ is explicitly written as 
\begin{align}
U_{12} = 
- c_{12} c_{23} s_{e} + e^{i \a} c_{13} c_{e} s_{12} + e^{i \b} s_{12}  s_{23} s_{13} s_{e} \, . 
\label{U12}
\end{align}
By substituting the solutions of $s_{23}$~(\ref{sol23}) and $s_{13}$~(\ref{sol13}) to  the constraint $|U_{12}|^{2} = C_{13}^{2} S_{12}^{2}$, $s_{12}^{2}$ is similarly obtained by solving a quadratic equation.
As in the case of $s_{13}$, the sign before the root sign can be absorbed into a term $2 c_{23} s_{e} (c_{13} c_{e} \cos \a + s_{13} s_{23} s_{e} \cos \b)$.

In particular, the dependence of $\b$ on $s_{12}$ is small, 
because it is suppressed by $s_{13} s_{e}$. 
Thus, the constraint on $s_{12}$ is approximately 
\begin{align}
C_{13}^{2} S_{12}^{2}  \simeq 
(c_{12} c_{23} s_{e})^{2} + (c_{13} c_{e} s_{12} )^{2}
- 2  c_{12} c_{23} s_{e} c_{13} c_{e} s_{12} \cos \a 
 \, . 
\label{c13s12}
\end{align}
Since this constraint has a similar structure to Eq.~(\ref{U132}), 
we choose $s_{12}^{2} \simeq A' + \cos \a \, B'   > 0$ as a positive solution that reproduces the matrix elements of $U_{\rm MNS}$. Here,  $A', B'$ are functions of several trigonometric functions.
We will not discuss the critical value of $s_{12}$ because it is relatively large for both the CKM and MNS matrices.

Imposing the three constraints, we can treat the CP violation in flavor-mixing matrices  as a function of $\a, \b
$ and $s_{e}$. Rough parameter dependences of the three solutions are as follows.
\begin{itemize}
\item $s_{23}$: 
From Eq.~(\ref{sol23}), dependences of $\a, \b$ on $s_{23}$ are
 relatively small because these phases depend on the second order of $s_{e}$ and $s_{13}$.
\begin{align}
s_{23} = S_{23} + O(s_{13}^{2}, s_{e} s_{13}, s_{e}^{2}) \, . 
\label{s23-1}
\end{align}

\item $s_{13}$: 
From Eq.~(\ref{U13}), $s_{13}$ depends on $\a - \b$. 
An expansion of $s_{13}$ for small $s_{e}$ is
\begin{align}
s_{13} = S_{13} + C_{13} S_{23} s_{e} \cos(\a - \b) + O(s_{e}^{2}) \, . 
\label{s13-1}
\end{align}
However, $s_{e}/ s_{13}$ is often not negligible, and this expansion does not have good  accuracy.

\item $s_{12}$: 
Since dependence of $\b$  is suppressed by $s_{13} s_{e}$ in Eq.~(\ref{U12}), 
$s_{12}$ has an almost only $\a$ dependence for a small $s_{e}$. 
 Therefore, the behavior of the solution is 
\begin{align}
s_{12} \simeq S_{12} + C_{12} C_{23} s_{e} \cos \a \, . 
\label{s12-1}
\end{align}

\end{itemize}
%

\section{The Jarlskog invariant and its numerical results}

In this section, we investigate numerical behaviors of the CP violation after imposing the three experimental constraints.
The Jarlskog invariant $J$ \cite{Jarlskog:1985ht} of the unitary matrix $U_{\rm MNS}$~(\ref{UVn2}) is
\begin{align}
J & = \Im [ U_{\m 2} U_{\t 3} U_{\m 3}^{*} U_{\t 2}^{*} ] \\
& = c_{13} c_{23} [
c_{12} s_{12} \left(  c_{13} s_{23} s_{13} (c_e^2-s_e^2) \sin \b  -c_{13}^2 s_{23}^2  c_e s_e \sin \a  \right ) 
 \nn \\
& + c_{23} (c_{12}^2 - s_{12}^{2}) s_{23} s_{13} c_e s_e \sin  (\a -\b )
+ s_{13}^2 c_e s_e \left(c_{23}^2 \sin \a  -s_{23}^2 \sin  (\a -2 \b ) \right) ] \, . 
\end{align}
To the extent that $s_{e}, s_{13} \lesssim 0.1$ holds, higher orders of $s_{e}$ and $s_{13}$ are negligible, and the dominant terms are those proportional to $\sin \a$ and $\sin \b$.
Note that the relative sign between terms depends only on $c_{e} s_{e}$.

From Eqs.~(\ref{s23-1}), (\ref{s13-1}), and (\ref{s12-1}), 
behavour of the Dirac phase $\d$ for small $s_{e}$ is
\begin{align}
\sin \d &= {J \over C_{12} S_{12} C_{23} S_{23} C_{13}^2 S_{13} } \\
& \simeq 
\sin \b  + s_{e} \lsp - {S_{23} \cos \b \sin (\a-\b) \over S_{13}}
+  { (1 - 2 S_{12}^2) C_{23} \cos \a \sin \b  \over C_{12} S_{12}} \rsp . 
\label{sd}
\end{align}
The term with $\sin (\a - \b)$ is a sum of the term with $\sin \a$ and a contribution from the expansion of $s_{13}$, while the term with $\cos \a$ is from the expansion of $s_{12}$.
For the MNS matrix, the second term is dominant over the third term because $S_{23} / S_{13} \simeq 5$ and $(1 - 2 S_{12}^2 ) C_{23}  / C_{12} S_{12} \simeq 1/2$ hold.
The same is true for the CKM matrix from $S_{23} / S_{13} \simeq 10$.

\subsection{Numerical results}

Plots of $\sin \d$ as a function of  $\a, \b$ and $s_{e}$ are shown in Fig.~2.
The range of $s_{e}$ is set within the critical value $|s_{e}| \leqq 0.2$. 
The green-shaded region corresponds to $\cos \d < 0$. 
From Eq.~(\ref{sd}), if $s_{e}$ is sufficiently small, $\d$ is approximately equal to $\b \sim \d_{\n}$. 
On the other hand, when $s_{e}$ approaches the critical value $s_{e}^{c}$, 
it shows singular behavour on the line $\a - \b = \pm \p/2, \pm 3 \p / 2$.
For $s_{e} > s_{e}^{c}$, the divided two regions correspond to the large and small solutions in Fig.~1.

\begin{figure}[t]
\begin{center}
 \includegraphics[width=16cm]{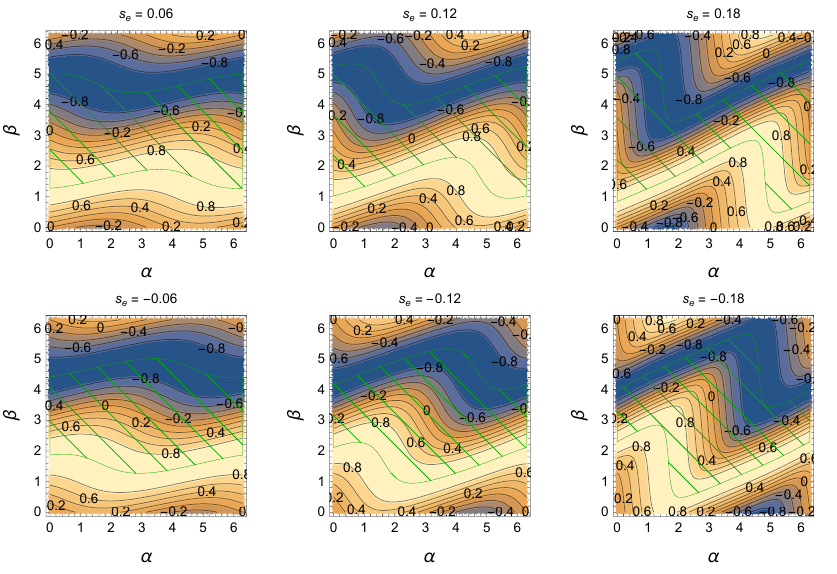}
\caption{Plots of $\sin \d$ of the MNS matrix $U_{\rm MNS}$ for a given 1-2 mixing $s_{e}$ and phases $\a, \b$. The green-shaded regions represent $\cos \d < 0$.
}
\end{center}
\label{}
\end{figure}

The DUNE and T2HK experiment has a discovery capability of around $\d \gtrsim 30^{\circ}$ after 10 years of operation.
This allows for physical interpretations when the next-generation experiments observe or do not observe the Dirac phase $\d$, respectively.
\begin{description}

\item{\bf When the Dirac phase (of about $O(1)$) is observed}

Large CP violation, as suggested by the T2K experiment, 
cannot be achieved by the relative phase $\a$ alone. 
Therefore, if the Yukawa matrix of charged leptons $Y_{e}$ is sufficiently hierarchical, 
large CP violation corresponds to the intrinsic CP violation $\d_{\n}$ in the diagonalization $U_{\n}^{0}$ of the neutrinos.
In particular, if the phase is close to $\pi/2$, the $\m - \t$ reflection symmetry \cite{Harrison:2002et, Grimus:2003yn, Grimus:2005jk} (and as a review \cite{Xing:2022uax}) is more favored.
However, in regions where $|s_{e} \sin \alpha|$ is large, the contribution of $\sin \beta$ can be subdominant.

\item{\bf When the Dirac phase (of about 30$^{\circ}$ or more) is does not observed}

Non-observation of CPV, as suggested by the NO$\n$A experiment, gives an upper bound $\sin \d^{\rm lim}$ on $\sin \d$.
In the absence of special cancellation, it imposes upper bounds on $\a \sim \r - \s $ and $\b \sim \d_{\n} $ respectively.
\begin{align}
{s_{e} S_{23} \over S_{13}} \sin \a \, ,  ~
\sin \b < \sin \d^{\rm lim} \sim 0.5 \, . 
\end{align}
In this situation, $\b = 0$ or $\pi$ can be favored, 
that is realized by some generalized CP symmetries \cite{ Yang:2020qsa, Yang:2020goc, Yang:2021smh, Yang:2021xob}. 
The smaller the observed Dirac phase, the more important the relative phase can be for the source of CPV. 
Besides, even for large $s_{e} \sim 0.1$, the relative phase $\sin \a$ is hardly restricted (and $\a \sim \pm \pi/2$ is allowed).

\end{description}

The reason for this physical interpretation is as follows: in Eq.~(\ref{UVn}), the mixing $s_{e}$ and $s_{\t}$ become zero when the chiral symmetry of the first and second generation becomes exact. 
In this limit, the phases $\r$ and $\s$ can be removed by further phase redefinition. Therefore, effects of the original phases $\r,\s$ are suppressed by $s_{e}$ and $s_{\t}$, and $\d_{\n}$ becomes the leading CPV source for small $s_{e}$.
Such restrictions on CP phases can provide important constraints on the leptogenesis scenario \cite{Fukugita:1986hr}.

\subsection{CKM matrix}

A similar analysis is performed for the CKM matrix. 
The 1-3 element of the CKM matrix satisfies 
\begin{align}
|V_{ub}| \sim m_{d1} / m_{d3} \gg m_{u1} / m_{u3} \, ,
\end{align}
and the approximation by chiral symmetry seems to be  reasonable.
As input parameters, we use the latest data of UTfit \cite{UTfit:2022hsi}. 
\begin{align}
\sin \th_{12}^{\rm CKM} &= 0.22519 \pm 0.00083 \, ,  ~~~ \sin \th_{23}^{\rm CKM} = 0.04200 \pm 0.00047 \, ,  \\
\sin \th_{13}^{\rm CKM} &= 0.003714 \pm 0.000092 \, ,  ~~~ \d_{\rm KM} = 1.137 \pm 0.022 \, . 
\end{align}

A similar plot for the CKM matrix is shown in Fig.~3. 
The parameter $s_{e}$ in the MNS matrix is redefined as $s_{u}$ and 
its range is set to be within the critical value ($|s_{u}| \leqq 0.09$). 
The green-shaded regions are excluded because $\cos \d_{\rm KM} < 0$.
The region within the error range of $\d_{\rm KM}$ is filled in. 
In the three continuous parameters, one two-dimensional surface is chosen that reproduces the observed CP phase.
\begin{figure}[t]
\begin{center}
 \includegraphics[width=16cm]{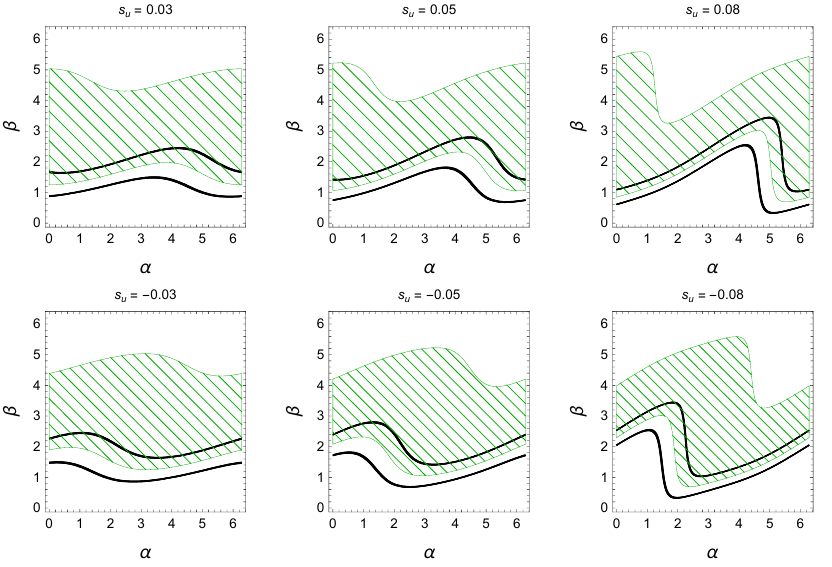}
\caption{ 
Plots of the CP violation $\sin \d_{\rm KM}$ of the quark mixing matrix $V_{\rm CKM}$. 
The green-shaded regions are excluded because they represent $\cos \d_{\rm KM} < 0$.
}
\end{center}
\label{}
\end{figure}

Since the phase behaves as $\d_{\rm KM} \simeq \b$ for sufficiently small $s_{u}$, 
we see that the observed CPV in the CKM matrix corresponds to the intrinsic CP phase $\d_{d}$ associated with the diagonalization of down quarks. 
This situation corresponds to the large CPV in the MNS matrix, and $|s_{u} \sin \alpha|$ must be large to make $\sin \beta$ small.
A characteristic structure of CP phases such as the $\m - \t$ reflection symmetry can be required.
On the other hand, near the critical value, 
a parameter set $s_{u} = 0.09, \a = -\pi/2$ or $s_{u} = -0.09, \a = \pi/2$ is possible. 
In the limit of $s_{13} \to 0$, Eq.~(\ref{UVn2}) is reduced to the Fritzsch--Xing parameterization \cite{Fritzsch:1997fw} and it has been known that the phase is almost maximal $|\a| = \pi/2$. 
Such restrictions can be useful in the construction of grand unified theories. 
 
\section{Summary}

In this letter, we perform an almost general analysis of flavor-mixing matrices $V_{\rm CKM}$ and $U_{\rm MNS}$ to investigate the discriminative power of CP phases by next-generation neutrino oscillation experiments.
For hierarchical Yukawa matrice $Y_{f}$ of the Standard Model fermions $f$, 
the unitary matrix $U_{f}$ of diagonalization for left-handed fields is expected to have small mixings. 
As an approximation, we neglect the 1-3 mixing of more hierarchical fermions (i.e. up-type quarks and charged leptons). 
Thus there are two sources of CP violation in $V_{\rm CKM}$ and $U_{\rm MNS}$, 
 the intrinsic CP phase $\d_{d, \n}$ in diagonalization of less hierarchical fermions $d, \n$ and relative phases between two unitary matrices. 

By eliminating unphysical phases and imposing constraints of the three measured mixing angles, 
the flavor-mixing matrices are analytically displayed by two phases and the 1-2 mixing $s_{u, e}$ of more hierarchical fermions.  
For sufficiently small 1-2 mixing $s_{e}$ of charged leptons, 
the Dirac phase $\d$ is mostly identical to the intrinsic phase of neutrinos $\d_{\n}$ 
and contributions of relative phases are not dominant.
Therefore, future detection of the Dirac phase indicates the observation of $\d_{\n}$. 
On the other hand, if such a CP violation is not observed, 
an upper limit is placed on a combination of $\d_{\n}$ and relative phases.
This interpretation holds because the relative phases can be eliminated by further phase redefinition 
in the limit of zero mixing of the charged leptons. 
Thus, contributions of relative phases are proportional to mixing angles of the charged leptons, 
which are suppressed by the chiral symmetries.

The analysis is also performed for the CKM matrix and similar conclusions are obtained.
Such restrictions will be useful in analyses of leptogenesis and grand unified theories.


\begin{thebibliography}{10}

\bibitem{T2K:2021xwb}
T2K collaboration, K.~Abe {\em et~al.},
\newblock Phys. Rev. D {\bf 103}, 112008 (2021), arXiv:2101.03779.

\bibitem{NOvA:2021nfi}
NOvA collaboration, M.~A. Acero {\em et~al.},
\newblock (2021), arXiv:2108.08219.

\bibitem{Hyper-KamiokandeProto-:2015xww}
Hyper-Kamiokande Proto-Collaboration, K.~Abe {\em et~al.},
\newblock PTEP {\bf 2015}, 053C02 (2015), arXiv:1502.05199.

\bibitem{DUNE:2020jqi}
DUNE Collaboration, B.~Abi {\em et~al.},
\newblock Eur. Phys. J. C {\bf 80}, 978 (2020), arXiv:2006.16043.

\bibitem{Hall:1993ni}
L.~J. Hall and A.~Rasin,
\newblock Phys. Lett. {\bf B315}, 164 (1993), arXiv:hep-ph/9303303.

\bibitem{Peccei:1995fg}
R.~D. Peccei and K.~Wang,
\newblock Phys. Rev. D {\bf 53}, 2712 (1996), arXiv:hep-ph/9509242.

\bibitem{Xing:2020ijf}
Z.-z. Xing,
\newblock Phys. Rept. {\bf 854}, 1 (2020), arXiv:1909.09610.

\bibitem{C:2014mmz}
S.~C, K.~N. Deepthi, and R.~Mohanta,
\newblock Adv. High Energy Phys. {\bf 2016}, 9139402 (2016), arXiv:1408.6071.

\bibitem{Ballett:2016daj}
P.~Ballett, S.~F. King, S.~Pascoli, N.~W. Prouse, and T.~Wang,
\newblock Phys. Rev. D {\bf 96}, 033003 (2017), arXiv:1612.07275.

\bibitem{deGouvea:2017yvn}
A.~de~Gouv\^ea and K.~J. Kelly,
\newblock Phys. Rev. D {\bf 96}, 095018 (2017), arXiv:1709.06090.

\bibitem{Kelly:2017kch}
K.~J. Kelly,
\newblock Phys. Rev. D {\bf 95}, 115009 (2017), arXiv:1703.00448.

\bibitem{Ge:2011qn}
S.-F. Ge, D.~A. Dicus, and W.~W. Repko,
\newblock Phys. Rev. Lett. {\bf 108}, 041801 (2012), arXiv:1108.0964.

\bibitem{Ge:2011ih}
S.-F. Ge, D.~A. Dicus, and W.~W. Repko,
\newblock Phys. Lett. B {\bf 702}, 220 (2011), arXiv:1104.0602.

\bibitem{Petcov:2014laa}
S.~T. Petcov,
\newblock Nucl. Phys. B {\bf 892}, 400 (2015), arXiv:1405.6006.

\bibitem{Girardi:2014faa}
I.~Girardi, S.~T. Petcov, and A.~V. Titov,
\newblock Nucl. Phys. B {\bf 894}, 733 (2015), arXiv:1410.8056.

\bibitem{Girardi:2015vha}
I.~Girardi, S.~T. Petcov, and A.~V. Titov,
\newblock Eur. Phys. J. C {\bf 75}, 345 (2015), arXiv:1504.00658.

\bibitem{Delgadillo:2018tza}
L.~A. Delgadillo, L.~L. Everett, R.~Ramos, and A.~J. Stuart,
\newblock Phys. Rev. D {\bf 97}, 095001 (2018), arXiv:1801.06377.

\bibitem{Fritzsch:1997fw}
H.~Fritzsch and Z.-Z. Xing,
\newblock Phys. Lett. {\bf B413}, 396 (1997), arXiv:hep-ph/9707215.

\bibitem{Gonzalez-Garcia:2021dve}
M.~C. Gonzalez-Garcia, M.~Maltoni, and T.~Schwetz,
\newblock Universe {\bf 7}, 459 (2021), arXiv:2111.03086.

\bibitem{Jarlskog:1985ht}
C.~Jarlskog,
\newblock Phys. Rev. Lett. {\bf 55}, 1039 (1985).

\bibitem{Harrison:2002et}
P.~F. Harrison and W.~G. Scott,
\newblock Phys. Lett. {\bf B547}, 219 (2002), arXiv:hep-ph/0210197.

\bibitem{Grimus:2003yn}
W.~Grimus and L.~Lavoura,
\newblock Phys. Lett. {\bf B579}, 113 (2004), arXiv:hep-ph/0305309.

\bibitem{Grimus:2005jk}
W.~Grimus, S.~Kaneko, L.~Lavoura, H.~Sawanaka, and M.~Tanimoto,
\newblock JHEP {\bf 01}, 110 (2006), arXiv:hep-ph/0510326.

\bibitem{Xing:2022uax}
Z.-z. Xing,
\newblock Rept. Prog. Phys. {\bf 86}, 076201 (2023), arXiv:2210.11922.

\bibitem{Yang:2020qsa}
M.~J.~S. Yang,
\newblock Phys. Lett. B {\bf 806}, 135483 (2020), arXiv:2002.09152.

\bibitem{Yang:2020goc}
M.~J.~S. Yang,
\newblock Chin. Phys. C {\bf 45}, 043103 (2021), arXiv:2003.11701.

\bibitem{Yang:2021smh}
M.~J.~S. Yang,
\newblock Nucl. Phys. B {\bf 972}, 115549 (2021), arXiv:2103.12289.

\bibitem{Yang:2021xob}
M.~J.~S. Yang,
\newblock PTEP {\bf 2022}, 013B12 (2021), arXiv:2104.12063.

\bibitem{Fukugita:1986hr}
M.~Fukugita and T.~Yanagida,
\newblock Phys. Lett. {\bf B174}, 45 (1986).

\bibitem{UTfit:2022hsi}
UTfit, M.~Bona {\em et~al.},
\newblock Rend. Lincei Sci. Fis. Nat. {\bf 34}, 37 (2023), arXiv:2212.03894.

\end{thebibliography}

\end{document}